\documentclass[twoside]{article}
\usepackage{fleqn,espcrc2}
\usepackage{graphicx}
\usepackage{here}

\newcommand{\AmS}{{\protect\the\textfont2
    A\kern-.1667em\lower.5ex\hbox{M}\kern-.125emS}}										
\def\beq{\vspace*{-0.2cm}\begin{equation}}
\def\eeq{\end{equation}\vspace*{-0.2cm}}
\def\bea{\vspace*{-0.2cm}\begin{eqnarray}}
\def\eea{\end{eqnarray}\vspace*{-0.2cm}}
\def\bq{\begin{quote}}
\def\eq{\end{quote}}

\def\nnb{\nonumber}
\def\ga{\left(}
\def\dr{\right)}

\def\rar{\rightarrow}
\def\lrar{\Longrightarrow}
																
\def\nnb{\nonumber}
\def\la{\langle}
\def\ra{\rangle}
\def\nin{\noindent}
\def\ba{\vspace*{-0.2cm}\begin{array}}
\def\ea{\end{array}\vspace*{-0.2cm}}

\def\b{$\bullet~$}
\def\als{\alpha_s}

\def\gg2{ \la\alpha_s G^2 \ra}
\def\gg3{g^3f_{abc}\la G^aG^bG^c \ra}
\def\ggg4{\la\als^2G^4\ra}

\def\calD{ {\cal D} }
\def\ftilde{\tilde f}

\def\therho{\theta\rho}

\title
{\bf{\boldmath
{\Large Glueball nature of the $\sigma/f_0(600)$ from  $\pi \pi$ and $\gamma\gamma$ scatterings } }}

\author{G. Mennessier\thanks{Email: gerard.mennessier@lpta.univ-montp2.fr} $^{\rm{a}}$, S. Narison\thanks{Email: snarison@yahoo.fr}  \address {\footnotesize Laboratoire
de Physique Th\'eorique et Astroparticules, Universit\'e
de Montpellier II, Case 070, Place Eug\`ene
Bataillon, 34095 - Montpellier Cedex 05,
France},
 and 
W. Ochs\thanks{Email: wwo@mppmu.mpg.de } \address {\footnotesize  Max-Planck Institut f\" ur Physik, 
D 80805 Munich, F\"ohringer Ring 6, Germany,} \\
}

\begin{document}

\pagestyle{myheadings}
\markright{ }
\begin{abstract}
\noindent
We estimate the $I=0$ scalar meson $\sigma/f_0(600)$ parameters
from $\pi\pi$ and $\gamma\gamma$ scattering data below 700 MeV using an improved analytic K-matrix model. 
A fit of the hadronic data gives a complex pole mass $M_\sigma= 422-i~290$ MeV, while simultaneous best fits of the $\gamma\gamma \to \pi^+\pi^-, \pi^0\pi^0$ data give a direct width of $ (0.13\pm 0.05)$ keV, a rescattering component of  $(2.7\pm 0.4)$ keV and a total (direct+rescattering) width of $(3.9\pm 0.6)$ keV.  ``Running" these results to the physical real axis, the small ``direct" $\gamma\gamma$ and the large hadronic widths at the ``on-shell" mass are compatible with QCD spectral sum rules (QSSR)  and some low-energy theorems (LET) expectations for an unmixed lowest mass glueball/gluonium $\sigma_B$ of a mass around 1 GeV and a large OZI-violation decay into $\pi\pi$ . 
\end{abstract}
\maketitle
\vspace*{-1.5cm}
\section{Introduction}
\vspace*{-0.25cm}
 \nin
Understanding the nature of scalar mesons in terms of quark and gluon
constituents
is a long standing puzzle in QCD \cite{MONTANET}. One might expect that
the decay rate of these mesons into two photons could provide an
 important 
information
about their intrinsic composite structure. The problem here is that
 some 
states 
are very broad
($\sigma$ and $\kappa$ mesons), others are close to an inelastic
 threshold
($f_0(980)$, $a_0(980)$), which makes their interpretation more
 difficult. 
Besides the interpretation within a $q\bar q$ model
\cite{MONTANET,MORGAN,BN,SNG,KLEMPT,OCHS,SN06}
or unitarized quark model \cite{TORN,BEVEREN},
 also the possibility of
tetraquark states \cite{JAFFE1,BLACK,LATORRE,SNA0,ACHASOV} (and some other related
 scenarios: meson-meson molecules \cite{ISGUR,BARNES}, 
meson exchange \cite{HOLINDE}) is considered. 
In addition, a gluonic meson is expected in
the scalar sector, according to lattice QCD \cite{PEARDON,MICHAEL}, 
QCD spectral sum rules (QSSR) 
\cite{NSVZ,CHET,SNG0,SNG1,VENEZIA,SNG,SN06} \`a la SVZ \cite{SVZ,SNB},
and some low-energy theorems (LET) \cite{VENEZIA,NOVIKOV,CHANO}. Such a
state could mix with the other $\bar qq$  mesons 
 \cite{BN,SNG,OCHS,CLOSE,aniso}. 
Among the light particles, the 
$f_0(600)/\sigma$ meson could be such a gluonic 
 resonance. It can manifest itself in some effective linear sigma models
 \cite{LANIK,SCHEC,ZHENG} or contribute to the low-energy constants at ${\cal O}(p^4)$ of
the QCD effective chiral Lagrangian \cite{PICH}.
Recent analyses of the $\gamma\gamma\to \pi\pi$ processes have extracted the
width of $f_0(600)/\sigma\to \gamma \gamma$ in the range: $ (4.1\pm 0.3) $
keV \cite{PENNINGTON},  $ (3.5\pm 0.7)$ or $(2.4\pm0.5) $ keV 
\cite{PENN1} to $(1.8\pm 0.4)$ keV \cite{OLLER}, while the one
from nucleon electromagnetic polarizabilities has given $(1.2\pm
0.4)~{\rm keV}$ \cite{PRADES}. 
Some of these results have been interpreted in \cite{PENNINGTON}  as disfavouring a
gluonic nature which is expected to have a small coupling to
$\gamma\gamma$
 \cite{VENEZIA,BN,SNG,ACHASOV,BARNES,SNA0}.
In the following, we shall reconsider the analysis  of the
 $\gamma\gamma\to \pi\pi$
process in the low energy region below 700 MeV, where we conclude that it
 is dominated by the coupling of the photons to charged pions and their
 rescattering, which  therefore can hide any direct coupling of the photons to
the scalar mesons. 
Some first results from this study have been presented in \cite{PETER}. 
\vspace*{-0.3cm}
\section{Results suggesting a light $0^{++}$ glueball}\vspace*{-0.25cm}
 \nin
The existence of glueballs/gluonia is a characteristic prediction of
 QCD and some
scenarios have been developed already back in 1975 \cite{MIN}. 
Today, there is agreement that such states exist in QCD and the
 lightest state has
quantum numbers $J^{PC}=0^{++}$. 

{\it Lattice QCD} - Calculations in the  simplified 
world without quark pair creation (quenched approximation) find the
lightest state at a mass around 1600 MeV \cite{PEARDON}. These findings
 lead to the construction of models
where the lightest glueball/gluonium mixes with other mesons in the
 nearby mass range
at around (1300-1800) MeV (see, for example, \cite{CLOSE}). 
However, recent results beyond this quenched approximation    
\cite{MICHAEL} suggest that the lightest state with a large gluonic
 component
is expected in the region around 1 GeV, and therefore, 
a scheme based on the mixing of meson states
with all masses higher than 1300 MeV could be insufficient to
represent the gluonic degrees of freedom in the meson spectrum.
Further studies concerning the dependence on lattice spacings and the quark
mass appear important.

{\it QSSR and LET} - 
These approaches have given quantitative
 estimates of the
mass of  glueballs/gluonia \cite{SNG0,SNG1,SNG} and of some essential
 features of its
branching ratios \cite{VENEZIA,SNG,SN06,NSVZ,LANIK}. As we shall see in section~\ref{qssr2},
these approaches require the existence of a gluonium $\sigma_B$ below 
1 GeV having a large
$\pi\pi$ but small $\gamma\gamma$ widths, in addition to its corresponding radial excitation $\sigma'_B $ and a narrow gluonium G(1.5-1.6) coupled weakly to $\pi\pi$ \cite{SNG0,VENEZIA,SNG,SN06}.

{\it Phenomenological studies} - 
The identification of the scalar glueball from experiment requires a full
understanding of the scalar meson spectrum. After grouping states into 
flavour multiplets the left over states are candidates for glueballs. 
Several schemes exist, motivated by the quenched lattice result, 
 where the extra gluonic state is assumed to mix into 
the three isoscalars $f_0(1370),\ f_0(1500)$ and $f_0(1710)$ \cite{CLOSE}. 
In an
alternative approach \cite{OCHS} (see also ref. \cite{ochs06}) the 
lightest $q\bar q$ multiplet is formed
from $f_0(980), f_0(1500), K^*_0(1430)$ and $a_0(1450)$, and the glueball
is identified as 
the broad object at smaller mass represented by $f_0(600$; 
it dominates $\pi\pi$ scattering near 1 GeV but extends from threshold up
to $\sim 1800$ MeV.
The appearance of this broad object 
in most gluon rich processes was considered 
in support of this hypothesis. In this analysis of the spectrum, 
results from elastic and
inelastic $\pi\pi$ scattering as well as from $D$, $B$ 
and $J/\psi$ decays have been considered \cite{OCHS,mo2,minkB}.

\vspace*{-0.3cm}
\section{Constraints on $\gamma\gamma$ from $\pi\pi$ processes}\vspace*{-0.25cm}
 \nin
Given a set of coupled multi-channel strong processes, 
 related electromagnetic or weak processes 
are largely
predetermined. First, there are the multi-channel extended unitarity relations
for the amplitudes with mixed strong and weak couplings
(corresponding to the Watson theorem for the single channel); secondly, there
are dispersion relations which these amplitudes have to fulfill (though not proved for composite particles).
The general consequences
of these constraints have been investigated by Omnes and Muskhelishvili
\cite{OMNES} for the single-
and multi-channel cases.
The amplitudes for the weakly coupled processes are largely determined by
strong amplitudes, but there are polynomial ambiguities which result from
subtraction terms of the dispersion relations depending on the asymptotic
high energy behaviour.
\vspace*{-0.3cm}
\subsection*{\b The analytic K-matrix model}
 \nin
This general formalism has been 
applied by Mennessier \cite{mennessier} (see also \cite{MENES2})
to the calculation of the electromagnetic
processes $\gamma\gamma\to \pi\pi,K\bar K$, given the strong processes
$\pi\pi\to \pi\pi,K\bar K$. The strong processes are represented by a K
matrix model representing the amplitudes by a set of resonance poles.
In that case, the dispersion relations in the multi-channel case 
can be solved explicitly, which is not possible otherwise. 
This  model can be reproduced by a set of Feynman diagrams, including 
resonance (bare) couplings to  $\pi\pi$ and $K\bar K$
 and 4-point $\pi\pi$ and $K\bar K$ interaction vertices.  
A subclass of bubble pion
 loop diagrams including resonance poles in the s-channel are resummed
 (unitarized Born).
The model also includes contributions from the exchange of vector mesons
in the t-channel which become important at the higher energies above 700
MeV.
The $\gamma\gamma$ scattering amplitudes also fulfill the constraints at
$s=0$ required by the Born approximation. 
The radiative width of the resonances cannot be predicted due to the polynomial ambiguity, which can be taken into account 
by introducing as free parameters the 
``direct couplings'' of the resonances to $\gamma\gamma$ in an effective
interaction vertex.
In the present analysis we have extended the model
by the introduction of a {\it shape function} which takes
explicitly into account left-handed cut singularities for the strong interaction amplitude.
This allows a more flexible parametrisation of the $\pi \pi$ 
data at low energies and 
improves the high energy behaviour.
Next we discuss the features of the model at low and at high energies.

\vspace*{-0.3cm}
\subsection*{\b Low energy limit: rescattering}
 \nin
A striking feature of the low energy $\gamma\gamma\to \pi\pi$ scattering is the
dominance of the charged over the neutral $\pi\pi$ cross section by an order
of magnitude which can be explained by
the contribution of the one-pion-exchange Born term in 
$\gamma\gamma\to\pi^+\pi^-$. In the process $\gamma\gamma\to
 \pi^0\pi^0$,
the photons cannot couple ``directly'' to $\pi^0\pi^0$ but through
 intermediate charged
pions and subsequent rescattering with charge exchange. 
This feature is realized in the analytic model \cite{mennessier}. A similar 
approach has been applied in
the model of Ref. \cite{ROSNER} which described data on $\gamma\gamma\to\pi
\pi$ up to 900 MeV by rescattering.
A modern example for the importance of rescattering 
is Chiral perturbation theory. To one-loop accuracy, the calculation based on one pion exchange and
$\pi\pi$ rescattering relates
the $\gamma\gamma\to \pi^0\pi^0$ to the $\pi^+\pi^-\to \pi^0\pi^0$
cross-section using
$m_\pi,~f_\pi$ and $e$ \cite{donoghue}.
 The  2-loop corrections obtained in \cite{bellucci} includes contributions
from other exchanges (spin 1). It amounts to about 30\% of the lowest 
order results and provides a reasonable description of the data
up to  $\sim700$ MeV. 
\vspace*{-0.3cm}
\subsection*{\b High energy limit: direct production}
 \nin
Whereas at low energy the photon interacts by its coupling to the charged
pions, the situation becomes different at high energies where the photon can
resolve the constituents of the hadrons. An example is the production of
$f_2(1270)$ in $\gamma\gamma\to \pi\pi$. The analytic model 
\cite{mennessier} predicts a
decreasing cross section for this process with pion exchange 
reaching $<10\%$ of the observed cross section at the peak of the $f_2$.
After inclusion of vector exchange, this contribution doubles. Because of
form factor effects involved in the photon coupling to hadrons, these
predictions from point like hadrons are to be taken rather as upper limits. 
This requires the need of ``direct coupling'' of $f_2\to \gamma\gamma$.
Indeed, it is well known that the radiative decays of the tensor mesons
$f_2,f_2',a_2$ are well described by a model with direct coupling to the
$q\bar q$ constituents according to the $SU(3)$ structure of the nonet 
with nearly ideal mixing (see, e.g. Ref. \cite{ARGUS}).  
Another example of a direct process is the production of $\rho$ mesons
in the process $\gamma p\to \pi^+\pi^-p$ by direct VMD coupling, 
which is found much larger
than the rescattering contribution from the ``Drell-S\"oding'' background
process \cite{SLACRHO}.  
We therefore interpret the direct terms in the $\gamma\gamma$ processes as
representing the amplitudes with coupling  of the photon to the partonic constituents
of the resonances (such as $q\bar q$, $4q$, $gg$,$\ldots$). 
\vspace*{-0.3cm}
\section{Direct  and rescattering couplings  of the $\sigma$}\vspace*{-0.25cm}
 \nin
In the application of the analytic model \cite{mennessier}, 
we first obtain a suitable
parametrisation of the $\pi\pi$ scattering data, and then determine the
direct coupling by comparing the model with the $\gamma\gamma$ results.
In the present Letter,
we restrict ourselves to the low mass region
below 700 MeV where we neglect vector and axial-vector exchanges~\footnote{They only start to be
relevant above 700 MeV \cite{mennessier}.}, 
inelastic channels and D-waves, furthermore, we assume a pointlike pion-photon
coupling. 
\vspace*{-0.3cm}
\subsection*{\b Amplitude for elastic $\pi\pi$ scattering}
 \nin
In the K-matrix fits to the $\pi\pi$ elastic scattering data for the range of energy [(0.5-1.8) GeV], a pole is found in the
isoscalar S-wave amplitude $  T_0^{(0)}$ with a large imaginary part which corresponds to a state 
\cite{CERN-MUNICH,ESTABROOKS,AMP}: 
\beq 
M\approx \Gamma\approx 1~{\rm GeV}, 
 \label{sigmamass} 
\eeq 
though this value may depend on the
treatment of the other resonances \cite{aniso,OCHS,ochs06}. 
The mass value in Eq. (\ref{sigmamass}) is close to the one where the 
observed S-wave phase shift goes through  $90^\circ$ as in a simple 
Breit Wigner form without background. This broad object has been called
$f_0(1000)$ \cite{AMP}, for a while $f_0(400-1200)$ by the PDG 
and now $f_0(600)$/$\sigma$.
In general, the complex resonance 
self-energy $\Sigma=M-(i/2) \Gamma$ 
is energy dependent
and determines the poles of the S-matrix corresponding to mass and width of
the physical particles; they do not usually coincide with the Breit Wigner
masses.
In recent determinations, where analyticity and unitarity properties 
are used to continue the amplitude into
the deep imaginary region,  one obtains for the complex $\sigma$ pole \footnote{A similar result [see Eq. (\ref{eq:pipiwidthfit})] is found in the analytic model
\cite{mennessier}.}:

\vspace*{-0.3cm}
\beq
 441-{\rm i}\ 272 ~{\rm MeV}~  \cite{leutwyler}\ \ {\rm or} \ \ 
489-{\rm i}~264~{\rm MeV}~  \cite{YND}~. \hfill 
\label{sigma}
\eeq

We shall see
that the $\sigma$  pole can be
referred to the same broad object defined above.
For applying the analytic model \cite{mennessier}, we introduce 
a {\it shape function} $f_0(s)$ which  multiplies the $\sigma\pi\pi$ coupling. 
For simplicity, we do not include the 4-point coupling
term.
The real analytic function $f_0(s)$ is regular for  $s > 0$
and has a left cut for  $s \le 0$. For our low energy approach, a convenient
approximation, which allows for a zero at $s=s_A$
and a pole at $\sigma_D>0$ simulating the left hand cut, is: 
\beq
f_0(s)=\frac{s-s_{A0}}{s+\sigma_{D0}} \label{formfactor}~.
\eeq 
The unitary 
$\pi\pi$ amplitude is then written as:
\beq
  T_0^{(0)}(s) = \frac{G f_0(s)}{s_R-s - G \ftilde (s)} = \frac{G f_0(s)}{\calD(s)}~.
\label{tpipi}
\eeq 
$T_\ell^{(I)}=e^{i\delta_\ell^{(I)}}\sin\delta_\ell^{(I)}/\rho$ with 
 $ \rho(s) =({1 - 4 m^2_\pi/s})^{1/2}$;  $G=g_{\pi,B}^2$ is the bare coupling squared 
 and:
\beq
{\rm {Im}}~ \calD = {\rm{Im}} ~ (- G \ftilde_0) = - (\therho) G \ f_0~,  
\label{eq:imaginary}
\eeq
with: $(\theta\rho)(s)=0$ below and $(\theta\rho)(s)=\rho(s)$
above threshold $s=4m_\pi^2$. 
The amplitude near the pole $s_0$ where $ {\cal D}(s_0)=0$ and
$\calD(s)\approx \calD'(s_0) (s-s_0)$ is:
\beq
  T_0^{(0)}(s)\sim \frac{g_\pi^2}{s_0-s}; \qquad g_\pi^2=\frac{G
f_0(s_0)}{-\calD'(s_0)}~.
\label{eq:gpi2}  
\eeq
The real part of $\cal D$ is obtained from a dispersion relation with
subtraction at $s=0$ and one obtains:
\beq
\ftilde_0(s) = \frac{2}{\pi} \Big{[} h_0(s) \ -h_0(0) \Big{]}~,
\label{eq:ftilde}
\eeq
where:
$h_0(s) =f_0(s) \tilde L_{s1}(s)$--$(\sigma_{N0}/(s+\sigma_{D0}))\tilde L_{s1}(-\sigma_{D0})$, 
$\sigma_{N0}$ is the residue of $f_0(s)$ at $-\sigma_{D0}$ and: $\tilde L_{s1}(s) =  
 \big{[}\ga{s - 4 m_\pi^2}\dr/{m_\pi^2} \big{]}
\tilde L_1(s,m_\pi^2)$ with $\tilde L_1$ from \cite{mennessier}. 
\vspace*{-0.3cm}
\subsection*{\b Amplitudes for $\gamma\gamma\to \pi\pi$ scattering}
 \nin
Starting from the
 S wave amplitude in Eq.~(\ref{tpipi}) we derive the amplitude $T_\gamma^{(I)}$ for 
the electromagnetic process for isospin $I=0$ as:
\beq
 T_{\rm \gamma}^{(0)} = 
  \sqrt{\frac{2}{3}} \alpha \left(f_0^B + G\ \frac{\ftilde_0^B}{\calD}\right) + 
  \alpha \frac{\cal P}{\calD}~. \label{elmamp} 
\eeq
Here the contribution from the Born term of $\gamma\gamma\to \pi^+\pi^-$
is given by $f_0^B=2L_1$ as defined
in \cite{mennessier}, a real analytic function in the $s$ plane 
with left cut $s\le 0$. The function 
$\ftilde_0^B$ represents $\pi\pi$ rescattering; 
it is regular for $s < 4 m_\pi^2$ but has a right cut 
for $s \ge 4 m_\pi^2$ with:
\beq
{\rm{Im}}~\ftilde_0^B (s+i\epsilon) =  (\therho \ f_0 \ f_0^B)(s)~,
\eeq
which vanishes at $s=0$.
With this definition the Watson theorem is fulfilled, i.e. the phase of 
$ T^{(0)}_\gamma$ is the same as the one of the elastic amplitude ${\cal D}^{-1}$
in Eq. (\ref{tpipi}). The real part is derived from a dispersion 
relation with subtraction at $s=0$ for $\ftilde_0^B$ to satisfy the Thomson limit,
and has a representation similar to the one in Eq. (\ref{eq:ftilde}), 
but by replacing
$(\tilde f_0,h_0)$ by $(\tilde f^B_0,h_0^B)$.
The function $h_0^B$ is defined as $h_0$ below Eq. (\ref{eq:ftilde}) but with 
$\tilde L_1$ replaced by $-\tilde L_1^2$ everywhere. It
vanishes at $s=-\sigma_{D0}$ and $ \ftilde_0^B(s)$
is regular at this point. 
Finally, the polynomial $\cal P$ reflects the ambiguity from the dispersion
relations and is set here to ${\cal P} = s F_\gamma\sqrt{2}$. It represents the direct
coupling of the resonance to $\gamma\gamma$. The residues 
 at the pole $s_0$ of the
rescattering  and direct contributions to $ T^{(0)}_\gamma$ in Eq.~(\ref{elmamp}), 
respectively, are obtained as: 
 \beq
g^{\rm resc}_\gamma  g_\pi =  \sqrt{\frac{2}{3}} \alpha 
 \frac{ G\ftilde_0^B(s_0)}{-\calD'(s_0)}; ~~~
    g^{\rm dir}_\gamma g_\pi = \alpha\frac{ s_0 F_\gamma \sqrt{2}}{-\calD'(s_0)},
 \label{eq:fgamma}
 \eeq
from which one can deduce the branching ratio:
\beq
{\Gamma_{\sigma\to\gamma\gamma}\over \Gamma_{\sigma\to\pi\pi}}
\simeq {1\over \vert \rho(s_0)\vert }\Big{\vert} {g_\gamma \over g_\pi
}\Big{\vert}^2
\simeq {2\alpha^2\over \vert \rho(s_0)\vert }\Big{\vert} {s_0\over Gf_0(s_0)}\Big{\vert} ^2 F_\gamma^2~.
\label{eq:gamwidth}
\eeq
Similarly, we parametrize
the  $I=2$ $S$-wave amplitude $T^{(2)}_0$ 
by introducing the shape function $f_2$:
\beq
 T^{(2)}_0= {\Lambda f_2(s) \over 1-\Lambda \tilde f_2(s)},
~~f_2(s)={s-s_{A2}\over (s + \sigma_{D1})(s + \sigma_{D2})}~,
\label{eq:t2f2}
\eeq
and obtain:
\beq 
T^{(2)}_\gamma={\alpha\over \sqrt{3}}\ga f_2^B+ {\Lambda \tilde f^B_2(s) \over 1-\Lambda \tilde f_2(s)}\dr~,
\label{eq:i2amplitude}
\eeq 
where $f_2^B=f^B_0$ and:
$
{\rm Im} \tilde{f_2}(s)  = (\theta  \rho) f_2(s),
~{\rm Im} \tilde{f^B_2}(s) = (\theta \rho~ f_2~ f_2^B)(s).
$
These amplitudes are again both subtracted at $s=0$ as in case of $I=0$ 
and one finds in analogy:
\beq
\tilde f_2(s) = \frac{2}{\pi}  \Big{[} h_2(s)-h_2(0)  \Big{]}~,
\label{f2tilfct}
\eeq
where:
$h_2(s) =
f_2(s)\tilde L_{s1}(s)-
({\sigma_{N1}}/({s+\sigma_{D1}}))
~\tilde L_{s1}(-\sigma_{D1})
 $ $  - ( {\sigma_{N2}}/({s+\sigma_{D2}}))\tilde L_{s1}(-\sigma_{D2})$;
$\sigma_{N1},\sigma_{N2}$ are the residues  of $f_2(s)$ at 
$-\sigma_{D1},-\sigma_{D2}$ and $\tilde f_2^B(s)$ is 
defined as $\tilde f_2(s)$ in Eq. (\ref{f2tilfct})
but with $\tilde L_{s1}$ replaced by  $-\tilde L_{s1}^2$.  
The cross sections for the $\pi\pi$ and $\gamma\gamma$ 
scattering processes are obtained 
from the previous expressions of the amplitudes
as described in the appendices of \cite{mennessier}, 
where the $\cos\theta$ integration should be done 
from 0 to 1 for the neutral and from -1 to 1 for the charged cases \footnote{The range of integration may have been misused in \cite{HARJES}.}.

\vspace*{-0.3cm}
\subsection*{\b Results of the analysis}
\begin{figure}[hbt]
\begin{center}
\includegraphics[width=8.cm]{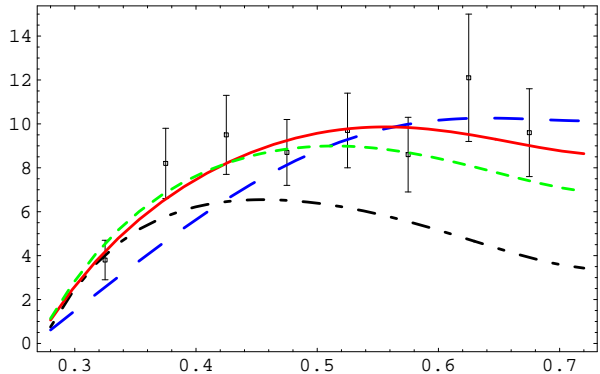}
\vspace*{-0.7cm}
\caption{\footnotesize   Fit of the $\pi^0\pi^0$ cross-section  in nanobarn (nb) versus $\sqrt{s}$ using unitarized Born amplitude: $F_\gamma=0$ (dot-dashed);
 $F_\gamma=-0.09$: I=0 (large dashed), I=0+2 (continuous); $F_\gamma=-0.07$: I=0+2 (small dashed). The data are from Crystal Ball \cite{cball} for $|\cos\theta| \leq 0.8$; 
 }
 \label{fig:fitneutral}
\end{center}
\vspace*{-1.cm}
\end{figure}

\begin{figure}[hbt]
\begin{center}
\includegraphics[width=8.cm]{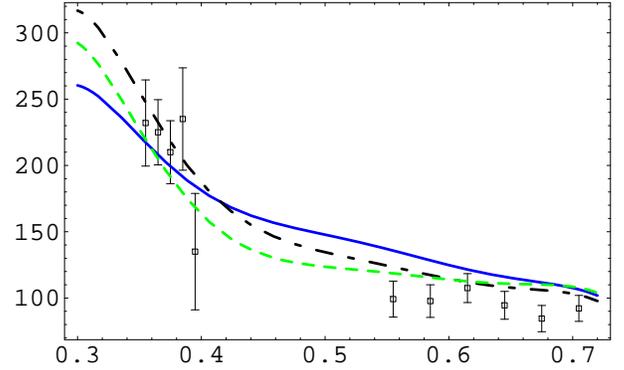}
\vspace*{-0.7cm}
\caption{\footnotesize  The same as in Fig. \ref{fig:fitneutral} but for $\pi^+\pi^-$ using unitarized Born amplitude: $F_\gamma=0$ (dot-dashed);
 $F_\gamma=-0.07$ and I=0+2 (small dashed). The continuous line corresponds
to the non-unitarized Born amplitude with $F_\gamma=0.$ 
The data are from MARKII \cite{MARK2} for $|\cos\theta| \leq 0.6$.}
\label{fig:fitcharged}
\end{center}
\vspace*{-1.cm}
\end{figure}

 \nin
The analytic model in its original parametrization  \cite{mennessier} 
without direct term
has given already quite a decent
description of the data available now for both charge states $\pi^+\pi^-$ and
$\pi^0\pi^0$ within $\sim 20\%$ as discussed previously \cite{PETER}. 
This already indicates  that the direct term is much smaller than the
rescattering term. In the
present analysis, we repeat the calculation but take the new version of the model \cite{mennessier}
by introducing the {\it shape function} $f_0(s)$ defined in Eq.~(\ref{formfactor}). 
We first determine the parameters of the model for $\pi\pi$ scattering below
700 MeV from the best approximation of our formula to the phase shifts
$\delta_0^{(0)}$ and $\delta^{(2)}_0$ obtained in the Roy equation analysis in \cite{leutwyler},
which takes into account the high-energy behaviour suggested in \cite{YND}. 
For $I=0$, we obtain in units of ${\rm GeV}^2$:
\vspace*{-0.3cm}
\bea
\vspace*{-0.25cm}
s_{A0}&=&0.016,~~~~~~~G\ \simeq  1.184~,\nnb\\
s_R&=&0.823,~~~~~~\sigma_{D0}= 0.501~,
\eea
and for $I=2$:
\vspace*{+0.25cm}
\bea 
s_{A2}&=&0.039,~~~~~~\Lambda\ \simeq  -4.136~,~\nnb\\
\sigma_{D1}&=&0.693,~~~~~~\sigma_{D2}= 4.384~,
\eea
from which, 
we find the pole mass and residue: 
\vspace*{+0.2cm}
\beq
 M_\sigma\simeq 422-{\rm i}~290~{\rm MeV} ~;~~g_\pi \simeq  0.06 -{\rm i}~0.50~{\rm GeV},
\label{eq:pipiwidthfit}\label{eq:gpi}
\eeq
which is close to the mass $M_\sigma$ in  Eq.~(\ref{sigma}) and in the 
previous paper  \cite{mennessier}. 
Given the $\pi\pi$ amplitude we can predict the cross sections for
$\gamma\gamma \to \pi\pi$  where the 
only free parameter $F_\gamma$ is related to the strength of the direct coupling 
$\sigma \to \gamma \gamma$. The fit of the model involving 
both the $\pi\pi$ rescattering and direct meson couplings to the data 
from Crystal Ball ($\pi^0\pi^0$) \cite{cball} and MARK-II ($\pi^+\pi^-$) \cite{MARK2}  collaborations
  is shown in 
Figs.~\ref{fig:fitneutral} and \ref{fig:fitcharged} for the cases with and without direct contribution.
 \vspace*{-.5cm}
A simultaneous best fit of the data in the region below 700 MeV, is obtained
for : 
$
F_\gamma \simeq - 0.070, 
$
at the minimum  $\chi^2 \simeq 25.6$ (19  data points)~\footnote{ In the earlier
analysis \cite{mennessier}, a value $F_\gamma$ of ${\cal O}(1)$ has been found,
which corresponds to the DM2 data in \cite{DM2} which are about twice the one of MARKII  \cite{MARK2}, i.e twice the Born contribution.}. 
However, a fit of  the neutral channel alone gives a much better fit: $
F_\gamma \simeq - 0.090
$, for a $\chi^2=3.38$ (8 data points).
In the charged channel, this is due 
to the deviations of the fit at high-mass; the large systematic errors 
and the absence of data points in the region 0.40 to 0.55 GeV 
 do not permit a good understanding of this channel.  
One can notice
that the $I=2$ $S$-wave amplitude which reproduces with high-accuracy 
the CGL points \cite{leutwyler}
improves the agreement with the data in the neutral case (Fig. \ref{fig:fitneutral}) but gives negligible contribution for the charged pion. 
For definiteness, we consider as a final estimate the mean value of the previous numbers:
\beq
F_\gamma \simeq - (0.080\pm 0.014)~, 
\label{eq:fcb} 
\eeq
where we consider that the uncertainty in the fitting procedure 0.010 is expressed by this range.  We have
added linearly  the systematics of the method to be about 5\% , which we have estimated from the deviation of the mass and width determinations in Eq. (\ref{eq:pipiwidthfit}) from the ones \cite{leutwyler} in Eq. (\ref{sigma}), as we have used their parametrization for fixing the parameters of the model.
The corresponding values of the residues in units of $ \alpha \times 10^{-3}$ GeV are:
\vspace*{-0.3cm}
\bea
g^{\rm dir}_\gamma  &\simeq& (7\pm 1)+{\rm i}~(31\pm 4)~,\nnb\\
g^{\rm resc}_\gamma  &\simeq& (90+5)+{\rm i}~(116\pm 6)~.
\eea
\vspace*{-0.3cm}
\\
Using Eqs. (\ref{eq:gamwidth}, \ref{eq:gpi}, \ref{eq:fcb}) and
$\Gamma_{\sigma\to \pi\pi}= 2 {\rm Im} M_\sigma$,  
we can deduce the ``partial" $\gamma\gamma$ widths at the complex pole :
\bea
\Gamma_{\sigma\to\gamma\gamma}^{\rm dir}&\simeq& (0.13\pm 0.05)~{\rm keV}~,\nnb\\
\Gamma_{\sigma\to\gamma\gamma}^{\rm resc}&\simeq& (2.7\pm 0.4)~{\rm keV}~,
\eea
 and the total $\gamma\gamma$ width (direct + rescattering):
 \vspace*{0.3cm}
 \beq
 \Gamma_{\sigma\to\gamma\gamma}^{\rm tot}\simeq (3.9\pm 0.6)~{\rm keV}~,
 \label{radtot}
\eeq
Improvements of these estimates need more precise data below 700 MeV, and an extension
of the analysis to higher energies.
\vspace*{-0.3cm}
\subsection*{\b Comparisons with existing results}

{\it  Direct $\gamma\gamma$ width} - Our result for the direct coupling is
comparable with our previous value \cite{PETER} obtained 
without using the shape function
$f_0(s)$.  In a parallel analysis \cite{ACHASOV07}, a K-matrix model
analogous to Ref. \cite{mennessier} has been applied in a larger energy
range of energy, including the recent BELLE data \cite{BELLE} and still
assuming elementary pion and kaon exchange; a smaller direct coupling was
obtained, but without error analysis.

{\it Total $\gamma\gamma$ width}
- Our result is compatible with the range of values
$(1.2 \sim 3.2)$ keV obtained in \cite{PETER} using a Breit-Wigner parametrization of the $\gamma\gamma\to \pi^0\pi^0$ data.
It is also  in the range of values based on dispersion relations (see introduction),
though a direct comparison is difficult to do as these analyses are extended to higher energy region,
and we have an additional determination of the direct term.
In our analysis below 700 MeV, we have also neglected the $\omega$ and
axial-vector exchange in agreement with \cite{mennessier}. Other
calculations find contributions of 2.5\% \cite{bellucci} or 16\%
\cite{OLLER} at 400 MeV.
One can notice that the effect of the direct term
in our Fig. \ref{fig:fitneutral}, distorts the shape of the unitarized Born
curve, and then permits a good fit of all points below 700 MeV, which is not
the case of the one in \cite{OLLER}. Then, if our fit was restricted to lower
masses to suppress these other exchanges no different 
results are to be expected.
\vspace*{-0.3cm}
\subsection*{\b ``On-shell" $\sigma$ mass and widths}
 \nin
For a more appropriate comparison with QSSR predictions ,
one can introduce the mass and widths of the ``visible meson" on the real axis.
This can be either the Breit-Wigner mass and width in Eq. (\ref{sigmamass}) or/and
the "on shell" mass and widths (see, e.g. \cite{SIRLIN}).

{\it Mass of the $\sigma$}
- A Breit-Wigner
parametrization of the data,  leads to value given in Eq. (\ref{sigmamass}):
$M_\sigma^{\rm bw}\approx \Gamma_\sigma^{\rm bw}\approx 1~{\rm GeV},
$
which can be compared with the ``on-shell mass" $M^{\rm os}_\sigma$ 
in the model \cite{mennessier} where the amplitude is purely imaginary 
at the phase 90$^0$: 
\beq
{\rm Re} {\cal D}({(M^{\rm os}_\sigma})^2)=0\lrar M_\sigma^{\rm os}\approx 0.92~{\rm GeV}~. 
\eeq

{\it Hadronic width }
- 
In the same way as for the mass, one can define an ``on-shell width"
 [see Eqs. (\ref{eq:imaginary}) and (\ref{eq:gpi2})] evaluated at $s=(M^{\rm os}_\sigma)^2$ :
\beq
M^{\rm os}_\sigma \Gamma^{\rm os}_\sigma\simeq {{\rm Im}~ \calD\over -{\rm Re~} {\cal D}'}
\lrar \Gamma_{\sigma\to\pi\pi}^{\rm os}\approx 1.02~{\rm GeV}~,
\eeq
to be compared with the Breit-Wigner width in Eq. (\ref{sigmamass}).

{\it Direct $\gamma\gamma$ width } 
- As there are no separated S wave
cross sections we use here only the estimates based on the analytic model 
\cite{mennessier} where we proceed as in the previous section 
but we evaluate the parameters at $s_0=(M^{\rm os}_\sigma)^2$. Then, we obtain from Eq.  (\ref{eq:gamwidth}) :
\beq
\Gamma_{\sigma\to \gamma\gamma}^{\rm os,dir}
\approx (1.0\pm 0.4)~{\rm keV}.
\label{eq:osgamwidth}
\eeq
As the model is extrapolated here towards
energies beyond its validity, we expect
that the previous results are a crude approximation
and will only serve as a guideline.

\vspace*{-0.3cm}
\section{QSSR/LET results and QCD tests of the $\sigma$} \label{qssr2}\vspace*{-0.25cm}
 \nin
The QSSR/LET 
results obtained in the physical region
can be better compared with experiment
by using the result for a Breit-Wigner parametrization of the data
in Eq. (\ref{sigmamass}) or by using the results for the on-shell mass derived previously but not
by directly comparing with the parameters of the complex pole. 
The direct $\gamma\gamma$ coupling, which can reveal the photon coupling to the intrinsic quark ($q\bar q,\ 4q,\ldots$)  or/and gluon ($\ gg,\ldots$)
constituent  internal  structure of the resonance, can be also related to the QSSR and/or LET evaluations of its width through quark or gluon loops like is the case of the quark triangle for the pion and/or the $f_2(1270)$. As the comparison with QSSR/LET is an important step for interpreting our fitted values, it is essential  to review shortly the main steps in the derivations of the results.
\vspace*{-0.3cm}
\subsection*{\b Gluonia masses from QSSR} 
 \nin
Masses of the bare unmixed scalar gluonium can be determined from the two Laplace unsubtracted (USR) and subtracted (SSR) sum rules:
\bea
{\cal L}_0(\tau)&=&{1\over\pi}\int_0^\infty dt e^{-t\tau}{\rm Im} \psi(t)~,\nnb\\
{\cal L}_{-1}(\tau)&=&-\psi(0)+{1\over\pi}\int_0^\infty {dt\over t} e^{-t\tau} {\rm Im} \psi(t)~,
\eea
\vspace*{0.2cm}
of the two-point correlator $\psi(q^2)$ associated to the trace of the QCD energy-momentum tensor current: 
\beq
\theta^\mu_\mu={1\over 4}\beta(\alpha_s) G^a_{\mu\nu}G^{\mu\nu}_a +\big{[} 1+\gamma_m(\alpha_s)\big{]}
\sum_{u,d,s} m_q\bar\psi_q\psi_q.
\eeq
$\tau$ is the sum rule variable; $\beta(\alpha_s)\equiv \beta_1(\alpha_s/\pi)+...$ and $\gamma_m(\alpha_s)\equiv \gamma_1(\alpha_s/\pi)+...$ are the QCD $\beta$-function and quark mass anomalous dimension: $-\beta_1=(1/2)(11-2n_f/3),~~\gamma_1=2$ for $SU(3)_c\times SU(n_f)$. The subtraction constant $\psi(0)=-16(\beta_1/\pi )\la \alpha_s G^2\ra$ expressed in terms of the gluon condensate \cite{NSVZ} $\la \alpha_s G^2\ra= (0.07\pm 0.01)~{\rm GeV}^4$ \cite{SNGLUE,SNB} affects strongly the USR analysis which has lead to apparent discrepancies in the previous literature when a single resonance is introduced into the spectral function \cite{SNG1}. The SSR being sensitive to the high-energy region $(\tau\simeq 0.3$ GeV$^{-2}$) predicts  \cite{SNG}:
\beq
M_G\simeq (1.5\sim 1.6)~{\rm GeV}~,
\eeq
comparable with the quenched lattice value \cite{PEARDON}, while the USR stabilizes at lower energy $(\tau\simeq 0.8$ GeV$^{-2}$) and predicts a low-mass gluonium \cite{SNG0}:
\beq
M_{\sigma_B}\simeq (0.95\sim 1.10) ~{\rm GeV}~,
\eeq
comparable with the unquenched lattice value \cite{MICHAEL}. 
Furthermore, the consistency
of the USR and SSR can be achieved by a two-resonances ($G$ and
$\sigma$) + ``QCD continuum" parametrization of the spectral function
\cite{VENEZIA}\footnote{In \cite{SNG} the QCD continuum has also been
modelized by a $\sigma'_B$ (radial excitation of the $\sigma_B$), which
enables to fix the decay constant $f_{\sigma_B,\sigma'_B}$ once the $\sigma_B,~\sigma'_B$
masses are introduced as input.}.
\vspace*{-0.3cm}
\subsection*{\b Gluonia couplings to Goldstone boson pairs}
 \nin
These couplings can be obtained from the vertex function:
\beq
V[q^2\equiv (q_1-q_2)^2]\equiv \la \pi_1\vert \theta_\mu^\mu\vert \pi_2\ra~,
 \eeq
 obeying a once subtracted dispersion relation:
 \beq
 V(q^2)=V(0)+q^2\int_{4m_\pi^2}^\infty {dt \over t}{1\over t-q^2-i\epsilon}{1\over \pi}{\rm Im} V(t)~.
 \eeq
with the condition:  $V(0)={\cal O}(m^2_\pi) \rar 0$ in the chiral limit. Using also the fact that $V'(0)=1$,  one can then derive the two sum rules:
\beq
{1\over 4}\sum_{S=\sigma_B,...} g_{S\pi\pi}\sqrt{2}f_S=0~,~~~
{1\over 4}\sum_{S=\sigma_B,...} g_{S\pi\pi}{\sqrt{2}f_S\over M_S^2}=1~,
\eeq
where $f_S$ is the decay constant analogue to $f_\pi$. The 1st sum rule requires the existence of at least two resonances coupled strongly to $\pi\pi$. Considering the $\sigma_B$ and $\sigma'_B$ but neglecting the small $G$-coupling to $\pi\pi$
as indicated by GAMS \cite{GAMS}\footnote{The $G(1.6)$ is expected
to  couple mainly to the $U(1)_A$ channels $\eta'\eta'$ and through mixing to $\eta\eta',~\eta\eta$ 
\cite{VENEZIA,SNG}.}, one predicts in the chiral limit \cite{VENEZIA,SNG} :
\beq
  |g_{\sigma_B\pi^+\pi^-}|\simeq 
    |g_{\sigma_B K^+K^-}| \simeq (4\sim 5)~{\rm GeV}~,
    \eeq
a universal coupling, which will imply a large width~\footnote {However, the analysis in Ref. \cite{SNG} also indicates  that $\sigma_B$ having a mass below 750 MeV cannot be wide ($\leq 200$ MeV) (see also some of Ref. \cite{SNG1})
due to the sensitivity of the coupling to $M_\sigma$. Wide low-mass gluonium has also been obtained
using QSSR (1st ref. in \cite{SNG1}), an effective Lagrangian \cite{LANIK} and LET \cite{NOVIKOV}.}:
\vspace*{-0.35cm}
\beq
\Gamma_{\sigma_B\to\pi^+\pi^-} \equiv  {|g_{\sigma_B
  \pi^+\pi^-}|^2\over {16\pi M_{\sigma_B}}}\ga 1-{4m^2_\pi \over
 M^2_{\sigma_B}}\dr^{1/2}\simeq0.7~ {\rm GeV}~.
  \label{eq:scalarwidth} 
\eeq
This large width into $\pi\pi$ is a typical OZI-violation  due to non-perturbative effects expected to be important in the region below 1 GeV, where perturbative arguments valid in the region of the G(1.5-1.6) 
cannot be applied. This result can be tested using lattice calculations with dynamical fermions.
\vspace*{-0.3cm}
\subsection*{\b Gluonia couplings to $\gamma\gamma$}
 \nin
These couplings can be derived by identifying the Euler-Heisenberg effective Lagrangian for  $gg\to \gamma\gamma$ via a quark
constituent  loop to the scalar one: $
{\cal L}_{S\gamma\gamma}= g_{S\gamma\gamma}SF_{\mu\nu}^{(1)}F_{\mu\nu}^{(2)},
$
 which leads to the sum rule \footnote{This sum rule has been used in \cite{NOVIKOV} for the charm quark.}:
\beq
g_{S\gamma\gamma} \simeq {\alpha\over 60}\sqrt{2}f_S M_S^2\ga{\pi\over-\beta_1}\dr \sum_{q=u,d,s}Q^2_q/M^4_q~, 
\eeq
where $Q_q$ is the quark charge in units of $e$; $M_{u,d}\approx M_\rho/2$ and   $M_{s}\approx M_\phi/2$;  are constituent masses; $S$ refers to gluonium ($\sigma_B,...)$. Then, one predicts the couplings:
\beq
g_{\sigma_B\gamma\gamma}\approx g_{\sigma'_B\gamma\gamma}\simeq g_{G\gamma\gamma} \simeq (0.4\sim 0.7)\alpha ~{\rm GeV}^{-1}~,
\eeq
and the corresponding widths \footnote{Due to their $M^3$-dependence, the  widths of the $\sigma'_B$ and $G$ can be much larger: $(0.4\sim 2)$ keV. These widths induce a tiny effect of $(3-9) \times 10^{-11}$ to the muon $(g-2)$ \cite{SNANOM} and cannot be excluded.} :
\beq
\Gamma_{\sigma_B\to\gamma\gamma}\equiv  {|g_{\sigma_B
\gamma\gamma}|^2\over 16\pi }M^3_{\sigma_B}\simeq (0.2\sim 0.6)~{\rm keV}~.
\eeq
 For a self-consistency check of the previous results, one can introduce their values into the sum rule:
  \beq
 {1\over 4}\sum_{S=\sigma_B,...} g_{S\gamma\gamma}\sqrt{2}f_S={\alpha R\over 3\pi}~,
 \eeq
 where $R\equiv 3\sum Q^2_q$, derived \cite{CHANO,LANIK} from  $ \la 0| \theta_\mu^\mu|\gamma_1\gamma_2 \ra $, by
 matching the $k^2$ dependence of its two sides.
 
  \vspace*{-0.1cm}
\subsection*{\b QCD tests of the $\sigma /f_0(600)$= a gluonium ?} 
 \nin
 One can notice that the QSSR and LET predictions for the mass, hadronic and 
electromagnetic widths of a low mass gluonium $\sigma_B$ 
are in remarkable agreement with previous results from 
  $\pi\pi$ and $\gamma\gamma$ scatterings for
 an on-shell or/and  Breit-Wigner resonance. However, the fitted value of 
direct width in Eq. \ref{eq:osgamwidth} to $\gamma\gamma$ needs to be improved for a sharper
comparison. 
 This ``overall agreement" favours a large gluon component in the $\sigma/f_0(600)$ wave function. 
 Its large width into $\pi\pi$ indicates a strong violation  of the OZI rule in this channel and 
 signals large non-perturbative effects in its treatment. This large hadronic width also disfavours its $\bar qq$ interpretation. In fact QSSR predicts for a $S_2\equiv (\bar uu+\bar dd)$, with mass of 1 GeV, a $\pi^+\pi^-$ width of about $(120\sim 180)$ MeV \cite{BN,SNG,SN06}\footnote{The 1st result in \cite{SNMENES} contains an unfortunate numerical error.}, and
a $S_2\to\gamma\gamma$ width of about  $5~{\rm keV}$. The later can be deduced
by using the quark charge squared relation (25/9) with $\Gamma_{a_0\to \gamma\gamma} \approx 2$ keV, obtained from the quark triangle loop including the $\la \bar\psi\psi \ra$ condensate contribution \cite{SNA0}.  In the same way, QSSR also  predicts, for a four-quark state, having the same mass of 1 GeV \cite{LATORRE,SNA0}, a $\gamma\gamma$ width of about 0.4 eV \cite{SNA0}. Both $\bar qq$ and 4-quark scenarios are
disfavoured by the value of the direct coupling fitted from $\gamma\gamma$ scattering.

\vspace*{-.35cm}
\section{Summary and conclusions}\vspace*{-0.25cm}
 \nin
We have reanalyzed  the $\pi\pi$ and $\gamma\gamma$ scattering data in the
mass region below 700 MeV.
We applied an analytic  $K$-matrix model which
is based on an effective theory with couplings between
resonances, hadrons and photons. The model around the threshold region resembles in its structure with Chiral
Perturbation Theory, while the inclusion of resonances and a resummation of
a class of meson loop corrections permit its applications towards higher energies. 
Then, the two-photon decay of a resonance
can proceed either through intermediate decay into hadrons and subsequent
annihilation or through the direct transition.
The low energy processes considered here
are dominated in our fits by the $f_0(600)/\sigma$ resonance. Its
total radiative width of about 4 keV is dominated by the rescattering component which is uniquely
predicted for known elastic  $\pi\pi$ scattering.
The direct coupling width is found an order of magnitude smaller ($0.13$
keV). It can be interpreted as representing its intrinsic quark ($q\bar q,\ 4q,\ldots$)  or/and gluon ($\ gg,\ldots$) constituent contribution to the $\gamma\gamma$ width.
\\
A confrontation of the overall fitted resonance parameters with the QSSR
predictions reviewed in Section \ref{qssr2}, for an unmixed bare gluonium $\sigma_B$ having a mass about 1GeV, indicate that the $\sigma/f_0(600)$ can
contain a large gluon component in its wave function. A light glueball/gluonium near
1 GeV is also suggested from lattice QCD calculations including dynamical fermions and from
the phenomenological analysis of the scalar meson spectrum.
\\
It will be interesting to extend this analysis towards higher energies
where new channels ($K\bar K$) will be opened and some other resonances
will show up. This allows for additional constraints on our approach  
and possible tests of the gluon/quark structure of $\sigma/f_0(600)$.
 \vspace*{-0.7cm}
\section*{Acknowledgements} 
\vspace*{-0.25cm}
\nin
We thank Peter Minkowski for the collaboration in an earlier
phase of this work.
One of us (W.O.) would like to thank Stephan
Narison for the invitation and for the kind hospitality at the Montpellier
University within the grant provided by Region of Languedoc-Roussillon
Septimanie where this work had begun. 
\vfill \eject

\end{document}